\documentclass[twocolumn]{aastex631}

\shorttitle{}
\shortauthors{Cohen et al.}

\graphicspath{{./}{figures/}}

\begin{document}

\title{Heating of the Atmospheres of Short-orbit Exoplanets by Their Rapid Orbital Motion Through an Extreme Space Environment}

\correspondingauthor{Ofer Cohen}
\email{ofer\_cohen@uml.edu}

\author[0000-0003-3721-0215]{Ofer Cohen}
\affiliation{Lowell Center for Space Science and Technology, University of Massachusetts Lowell, 600 Suffolk Street, Lowell, MA 01854, USA}

\author[0000-0001-9843-9094]{Alex Glocer}
\affiliation{NASA Goddard Space Flight Center, Greenbelt, MD, USA }

\author[0000-0002-8791-6286]{Cecilia Garraffo}
\affiliation{Harvard-Smithsonian Center for Astrophysics, 60 Garden Street, Cambridge, MA 02138, USA}

\author[0000-0001-5052-3473]{Juli\'{a}n D. Alvarado-G\'{o}mez}
\affiliation{Leibniz Institute for Astrophysics Potsdam, An der Sternwarte 16, 14482 Potsdam, Germany}

\author[0000-0002-0210-2276]{Jeremy J. Drake}
\affiliation{Lockheed Martin, 3251 Hanover St., Palo Alto, CA 94304, USA}

\author[0000-0002-5688-6790]{Kristina Monsch}
\affiliation{Harvard-Smithsonian Center for Astrophysics, 60 Garden Street, Cambridge, MA 02138, USA}

\author{Farah Fauth Puigdomenech}
\affiliation{University of Southampton, Southampton, SO17 1BJ, United Kingdom}


\begin{abstract}

Exoplanets with short orbit period reside very close to their host stars. They transition very rapidly between different sectors of the circumstellar space environment along their orbit, leading to large variations of the magnetic field in the vicinity of the planet on short timescales. This rapid change of the magnetic flux through the conducting and resistive layer of the planetary upper atmosphere may drive currents that dissipate in the form of Joule Heating. Here, we estimate the amount of Joule Heating dissipation in the upper atmosphere of Trappist-1e, and two hypothetical planets orbiting the Sun in close-in orbits. We find that the rapid orbital motion could drive a significant amount of atmospheric heating and could significantly affect the planetary atmosphere escape rate. Thus, the process should be accounted for when studying the long-term evolution of exoplanetary atmospheres.

\end{abstract}

\keywords{}


\section{Introduction} \label{sec:intro}

The upper atmosphere of planets is the outermost layer which directly interacts with its nearby space environment created by their host star (assuming the planet has an atmosphere). In the case of the Earth, this interaction is known as Space Weather which drives geomagnetic activity \citep{Kivelson.Russell:95,Heliophysics09,SpacePhysics16}. Any planet with even a thin atmosphere (such as Mars) would have its upper atmosphere ionized by the stellar ionizing Extreme Ultra Violet (EUV) radiation, creating a conducting layer at the top of the atmosphere through which the magnetospheric currents close --- the Ionosphere. Since the electrical conductivity in the ionosphere is finite, ionospheric currents dissipate, leading to a Joule Heating (JH) of the upper atmosphere. In the case of the Earth, JH energy flux is of the order of $10^{-2}~Wm^{-2}$ \citep{2011JGRA..116.5313D}.

Short-orbit exoplanets, with a few days orbital period, reside extremely close to their host stars. At these orbits, both the stellar EUV radiation, as well as the stellar wind conditions are orders of magnitude higher than the conditions around the Earth. As a result, a large amount of energy is transferred from the space environment onto the planet. Extreme EUV radiation could lead to a heating of the atmosphere, and the driving of hydrodynamic atmospheric escape  \citep[photoevaporation, see e.g.,][]{watson_dynamics_1981,Lammer2003,Tian2005extra,Schneiter2007,Penz2008,Murray-Clay2009,Owen2012,Tripathi2015,salz2016energy,Owen19}. Similar atmospheric heating could also come from an internal source \citep[Core-power escape,][]{Ginzburg2018}.  

JH could also be generated by the dissipation of ionospheric currents, which are driven by the stellar wind motional electric field \citep[Direct Current or DC system, e.g.,][]{Kivelson.Russell:95}. \cite{2018ApJ...856L..11C} showed that significant JH of the upper atmosphere of three of the Trappist-1 exoplanets is possible due to the extreme stellar wind conditions. They used the formalism presented by \cite{KivelsonRidley:08}, which introduces the combination of stellar wind density and magnetic field as an Alfv\'en wave, which is characterized by an Alfv\'enic "impedance". In this formalism, the amount of JH of the upper atmosphere is determined by the relation between that stellar wind Alfv\'enic impedance and the ionospheric impedance. \cite{2018ApJ...856L..11C} showed that the energy flux associated with the JH due to the extreme stellar wind conditions could be larger than the EUV energy flux in the Trappist-1 planets, and it could reach a few percent of the total stellar constant energy flux. Similar trends were obtained by \cite{2014ApJ...796...16K}, who calculated the JH in the atmospheres of in tidally-locked hot jupiters due to strong zonal flows \citep[e.g.,][]{2010exop.book..471S,2010ApJ...714L.238B,2014ApJ...782L...4R,2014ApJ...796...16K}.

In short-orbit exoplanets, strong variations of the gravitational forces due to the orbital motion could lead to a strong tidal heating of the planet's interior \citep[e.g.,][]{2018ApJ...857..106V}. As the planet orbits its host star, it is rapidly transitioning between different space environment sectors that have a different value for the ambient magnetic field, $B$. This strong $dB/dt$ may drive a voltage and currents that will dissipate in the ionosphere of the exoplanet in the form of Ohmic dissipation. This is a Voltage Driven JH (VDJH, Alternate Current or AC system), which is driven by the planetary motion itself. \cite{2017NatAs...1..878K} have studied the heating of the interiors of the Trappist-1 planets due to this effect. However, atmospheric heating has not been investigated yet.

The study presented here aims to estimate the importance of the VDJH in short-orbit exoplanets. In Section~\ref{sec:model}, we formulate a framework and the geometry of the problem, which enable us to estimate the amount of energy flux that goes into the planetary atmosphere due to VDJH. In Section~\ref{sec:results}, we present the results for a set of assumed parameters and we compare the VDJH energy flux to the energy fluxes due to EUV radiation and the stellar constant radiation, which are the dominant energy inputs from the stellar environment into planetary atmospheres. We discuss the results in Section~\ref{sec:discussion}, and we conclude our findings in Section~\ref{sec:summary}.


\section{Model} \label{sec:model}

\subsection{General Formalism}

We formalize our theoretical model for the possible heating of the planetary upper atmosphere of short-orbit exoplanets under the following underlaying assumptions. First, we assume that due to the short orbit and the fast orbital motion of the planet, the variations in the background stellar wind magnetic field are also rather fast. Second, we neglect the possible existence of a planetary internal field, $B_p$, since the change in magnetic flux through the upper atmosphere should not be affected by the internal planetary field, only the external, stellar wind field ($dB_p/dt=0$). We also neglect any field variations created by induced currents in the ionosphere. Finally, we assume a constant conductivity across the ionosphere. The last two assumptions are of course an approximation that needs a further study. Nevertheless, our study here aims to provide some upper/lower limits for the effect.

We start with the integral form of Faraday's law, which states that a time-varying magnetic flux, $\Phi$, generates an Electromotive Force, $\mathcal{E}$ (EMF, a voltage generator):
\begin{equation}
\mathcal{E} = -\frac{d \Phi}{dt},
\label{eq1}
\end{equation}
where the total magnetic flux, $\Phi$, is defined as 
\begin{equation}
\Phi=\int \mathbf{B}\cdot\mathbf{ds},
\label{eq2}
\end{equation}
with $\mathbf{B}$ being the magnetic field and $\mathbf{ds}$ being a surface element. For a constant area, $A$, the magnetic flux is simply $\Phi=A B$, where $B$ here is the magnetic field which crosses the area $A$. For a constant area, Eq.~\ref{eq1} becomes:
\begin{equation}
-\frac{d\Phi}{dt}=-A\frac{d \mathbf{B}}{dt} = \mathcal{E}=\mathbf{I}R,
\label{eq3}
\end{equation}
where $\mathbf{I}$ is the induced electric current driven by the EMF, and $R$ is the electrical resistance. The electrical resistance can be related to the electrical conductivity, $\sigma$:
\begin{equation}
R=\frac{l}{\sigma a}.
\label{eq4}
\end{equation}
We note that this conductivity has SI units of $[S/m]$, and it is different from the height-integrated conductivity, with units of $[S]$. $a$ is the cross section of the conductor/resistor through which the current $\mathbf{I}$ is flowing and $l$ is the length of the current loop. With these definitions, we can express the total current, $\mathbf{I}$, in terms of the current density, $\mathbf{j}$, and the cross section, $a$:
\begin{equation}
\mathbf{I}=a\mathbf{j},
\label{eq5}
\end{equation}
so Eq.~\ref{eq3} becomes:
\begin{equation}
-A\frac{d \mathbf{B}}{dt} = \mathbf{I}R=a\mathbf{j}\frac{l}{a\sigma}=\frac{\mathbf{j}l}{\sigma},
\label{eq6}
\end{equation}
and we can express the current density in terms of the rate of change of the ambient magnetic field:
\begin{equation}
\mathbf{j}=-\frac{A\sigma}{l}\frac{d \mathbf{B}}{dt}.
\label{eq7}
\end{equation}

\begin{figure*}[h!]
\centering
\includegraphics[width=5.in]{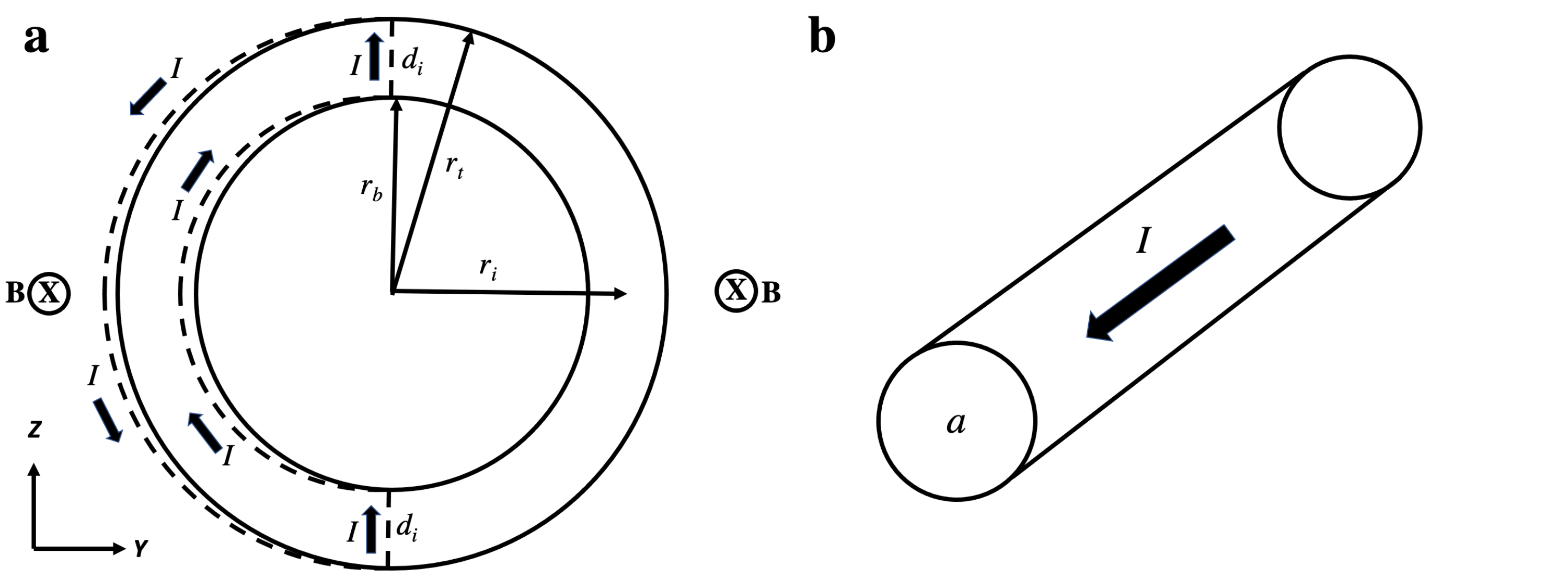}
\caption{Left: the geometry of the problem. The stellar wind magnetic flux is going through the ionospheric cross section, which is determined by the area bounded between $r_b$ and $r_t$. The current, $I$, induced by $dB/dt$ is flowing in a closed loop with an approximated length $2\pi r_i$ (dashed line, see Eq.~\ref{eq9}) and a cross section, $a$ (shown on the right).}
\label{fig:SystemDiagram}
\end{figure*}

Figure~\ref{fig:SystemDiagram} shows the geometry of the problem we formulate here. We assume a short-orbit exoplanet that sweeps through the interplanetary medium and the ambient stellar wind and that the ambient stellar wind magnetic field is radial in the frame of reference of the star due to the stretching by the stellar wind plasma. We neglect the azimuthal component of the ambient field since both the radial and the azimuthal field should contribute to the magnetic flux perpendicular to the ionosphere cross section. If a strong azimuthal field exists, it would only matter for the radial-azimuthal partition but not for the time-variations of the field magnitude. The star-planet line is defined as the X-direction (the planetary day side is facing the star) so that the ambient magnetic field is perpendicular the the planetary day side cross section, which is in the Y-Z plane. We focus our problem on the narrow conducting spherical layer at the top of the planetary atmosphere --- the ionosphere, which is bounded by the ionosphere bottom, $r_b$, and the ionosphere top, $r_t$. The average altitude of the ionosphere's center is $r_i=(r_b+r_t)/2$, and the thickness of the ionosphere is $d_i=r_t-r_b$.

\subsection{Shallow Ionosphere Limit}

In the limit where the ionosphere thickness is much smaller than the average height ($r_i>>d_i$), and assuming that the ambient magnetic flux of the stellar wind is going through a spherical cross section bounded between $r_b$ and $r_t$, the cross section can be approximated as:
\begin{equation}
A = \int^{r_t}_{r_b} 2\pi r dr \approx 2\pi r_i d_i.
\label{eq8}
\end{equation}

The change in the magnetic flux will drive a current in the ionosphere that will act to reduce that change (hence the minus sign in eq.~\ref{eq1}). Ideally, the induced current flows in a closed loop which encloses the area through which the changing magnetic flux flows. In the more realistic scenario of the ionosphere, the current will flow through the path of highest conductivity, which will be defined by the local, non-uniform distribution of the ionospheric conductivity. As a lower limit for our problem, we consider the case where the current path, $l$, encloses half of the ionosphere cross section. This path is shown as dashed line in Figure~\ref{fig:SystemDiagram}  a. Since $j$ is inversely proportional to $l$, this path provides the lower limit value for $j$ , as well for the total JH. Using this path, we have:
\begin{eqnarray}
&l=\frac{1}{2}2\pi r_b+\frac{1}{2}2\pi r_t+2d_i=2\pi\frac{r_b+r_t}{2}+2d_i=& \\
&2\pi r_i+2d_i\approx 2\pi r_i& \nonumber
\label{eq9}
\end{eqnarray}. 
With these assumptions, and taking half of the cross section, $A$, Eq.~\ref{eq7} becomes:
\begin{equation}
\mathbf{j}=-\frac{1}{2}\frac{2\pi r_i d_i\sigma}{2\pi r_i}\frac{d \mathbf{B}}{dt}=-\frac{d_i \sigma}{2} \frac{d \mathbf{B}}{dt}.
\label{eq10}
\end{equation}
Eq~\ref{eq10} can be written in the form of Ohm's law:
\begin{equation}
\mathbf{j}=\sigma \mathbf{E},
\label{eq11}
\end{equation}
with $\mathbf{E}=-\frac{d_i}{2} \frac{d \mathbf{B}}{dt}$. The Ohmic dissipation (JH), $q$, is given by $q=\mathbf{j}\cdot \mathbf{E}=j^2/\sigma$, or:
\begin{equation}
q=\frac{j^2}{\sigma}=\frac{d^2_i\sigma^2  \left( \frac{d B}{dt} \right)^2}{4\sigma}=\frac{d^2_i\sigma}{4}  \left( \frac{d B}{dt} \right)^2
\label{eq12}
\end{equation}
Thus, the volumetric power of the JH as a function of the change in the ambient magnetic field is:
\begin{equation}
q= \frac{d^2_i \sigma}{4} \left( \frac{d B}{dt} \right)^2~[W~m^{-3}].
\label{eq13}
\end{equation}
The energy flux, $Q$, associated with the JH is given by multiplying Eq.~\ref{eq13} by the width of the ionosphere, $d_i$:
\begin{equation}
Q=qd_i= \frac{d^3_i \sigma}{4} \left( \frac{d B}{dt} \right)^2~[W~m^{-2}],
\label{eq14}
\end{equation}
which is the total energy flux that is transferred to the ionosphere due to the change of the ambient stellar wind magnetic field. 

\subsection{Extended Ionosphere Limit}

In the limit of a very extended atmosphere, $d_i \approx r_t$. In this case, the ionosphere cross section may be approximated as: 
\begin{equation}
A = \int^{r_t}_{r_b} 2\pi r dr =\pi(r^2_t-r^2_b)\approx \pi r^2_t\approx \pi d^2_i,
\label{}
\end{equation}
and the length of the current loop can be approximated as: 
\begin{equation}
l =\frac{1}{2}2\pi d_i+2d_i \approx 5d_i.
\end{equation}
With these approximations, the current density becomes:
\begin{equation}
\mathbf{j}=-\frac{\pi d^2_i}{2}\frac{\sigma}{5d_i}\frac{d \mathbf{B}}{dt} \approx -0.3d_i \sigma \frac{d \mathbf{B}}{dt}.
\end{equation}
In this limit, the energy flux is:
\begin{equation}
Q \approx  0.3^2d^3_i \sigma \left( \frac{d B}{dt} \right)^2=0.1d^3_i \sigma \left( \frac{d B}{dt} \right)^2~[W~m^{-2}],
\end{equation}
which is about half of the heating in the case where $d_i<<r_i$.

In a real ionosphere, the current does not flow in a wire-like, one-dimensional manner, but it flows in an extended region. The current, as well as its dissipated JH drop from a central region since the current acts to cancel the change in the magnetic flux. The size of the drop can be characterized by a "skin depth", $\delta$, which can be estimated using Eq. 5.165 from \cite{1998clel.book.....J}:
\begin{equation}
\delta=\sqrt{\frac{2}{\sigma \mu \omega}},
\end{equation}
with $\omega$ being the characteristic frequency of the time-varying magnetic field, and $\mu$ being the magnetic permeability of the material. Here we assume that $\mu=\mu_0$, the permeability of free space. If we assume that the dominant period, $T$, is of the order of one day (based on values shown in Section~\ref{sec:results}), then for $\omega=2\pi/T$ and $\sigma$ values of 0.1 and 10 $Sm^{-1}$, we get a skin depth in the range of 100-1000 km. This means that the JH is not extremely local, but it is expanding from the skin around the ionospheric cross section, where the current flows. A more detailed estimate about the overall transfer of the VDJH requires a more detailed modeling effort.

\subsection{Stellar Wind Model}

VDJH depends on the variations of the Interplanetary Magnetic Field (IMF) strength along the planetary orbit. Such detailed IMF data is not available for exoplanets \citep[some observations were made for the stellar wind interaction with the interstellar medium, e.g.,][]{2021ApJ...915...37W}, nor it is available for short-orbits around the Sun \citep[limited data at specific locations is available from the Parker Solar Probe,][]{2023SSRv..219....8R}. Due to the lack of observational constrains, we must rely on models to estimate the relevant stellar wind conditions. Physics-based models for the solar corona and the solar wind have been vastly validated against observations. Thus, they can provide the most reliable tool to study stellar winds of solar analogs, which are not observationally constrained. Such a model is the Alfv\'en Wave Solar Model \citep[AWSoM,][]{2014ApJ...782...81V}, which is based on the BATSRUS Magnetohydrodynamic (MHD) model \citep{1999JCoPh.154..284P,2012JCoPh.231..870T}. The model has been vastly used to study the stellar winds of Sun-like stars in the context of the space environment around exoplanets \citep[e.g.,][]{2014ApJ...790...57C,2016ApJ...833L...4G,2017ApJ...843L..33G,2018PNAS..115..260D,2018MNRAS.481.5296V,2019ApJ...875L..12A,2020ApJ...897..101C,2022ApJ...928..147A}.

The model provides a three-dimensional numerical solutions for the solar/stellar wind from the stellar coronal base to a few tenths of stellar radii (depending on the specific star).  The model is driven by the distribution of the photospheric radial magnetic field of the star, which is imposed as the inner boundary condition. These synoptic maps, or magnetograms, are obtained for the Sun using e.g., the Michelson Doppler Imager \citep[MDI,][]{1991AdSpR..11d.113S} and the Helioseismic Magnetic Imager \citep[HMI,][]{2007AN....328..339K}. For stars, low-resolution magnetic maps are obtained using the Zeemann Doppler Imaging (ZDI) technique \citep{1989A&A...225..467D}. Using the observed boundary conditions, an analytic three-dimensional solution for the magnetic field is calculated using the potential field method \citep{1969SoPh....9..131A}. This three-dimensional magnetic field is served as the initial condition in the model.

Once the initial, static (potential) conditions are defined, the model solves the set of non-ideal MHD equations which includes the conservation of mass, momentum, magnetic induction, and energy. The momentum equation includes an Alfv\'en wave pressure gradient term, which accelerates the solar wind, while the energy equation includes heating and cooling terms that depends on the Alfv\'en wave heating, electron heat conduction, and the radiative cooling. The model is run until an MHD, forced steady-state is obtained in the three-dimensional simulation domain. We refer the reader to \cite{2014ApJ...782...81V} and \cite{2021ApJ...908..172S} for a complete description of the model. From this three-dimensional solution, simulated data can be extracted along a selected trajectory to provide the stellar wind conditions and magnetic field variations along a planetary orbit.  

\subsection{Estimated orbital magnetic field variations} 

For our investigation here, we simulate the solar wind conditions during Carrington Rotation 1916, which represents solar minimum conditions (the model was driven by an MDI magnetogram). In the context of our investigation, the solar wind conditions should not vary much from solar minimum to solar maximum as only the wind's spatial distribution would change due to the change and reversal of the solar magnetic field. We extract this solar wind solution along two circular orbits close to the Sun - one at 0.05 Astronomical Units (AU), and one at 0.1 AU. These orbits represent the orbits of many known short-orbit exoplanets (e.g. the NASA exoplanetary encyclopedia\footnote{https://exoplanets.nasa.gov/discovery/exoplanet-catalog/}). Thus, these orbital solutions represent typical variations of the IMF along the orbits of many exoplanets orbiting G -type stars in close-in orbits. For a case of an M-dwarf exoplanet, we use the stellar wind conditions along the orbit of the planet Trappist-1e \citep{2016Natur.533..221G} as calculated by \cite{2017ApJ...843L..33G}. While ZDI data is not available for Trappist-1, \cite{2017ApJ...843L..33G} have used ZDI data of the proxy star GJ 3622. It has been demonstrated that stars with available ZDI data could be used as proxies for stars with a similar Rossby number and without available ZDI data \citep{2016ApJ...833L...4G,2017ApJ...843L..33G,2022ApJ...941L...8G}. 

Our goal in this paper is to demonstrate that VDJH could be significant in some exoplanets and it should be taken into account. The three synthetic datasets for $B$, as well as $dB/dt$ allow us to calculate the VDJH in a general manner, one based on the estimated conditions around an M-dwarf star, and two that are based on the Sun itself (a G-type star). This range of conditions captures most of the possible intermediate exoplanet cases orbiting Sun-like stars. 

\begin{figure*}[h!]
\centering
\includegraphics[width=6.in]{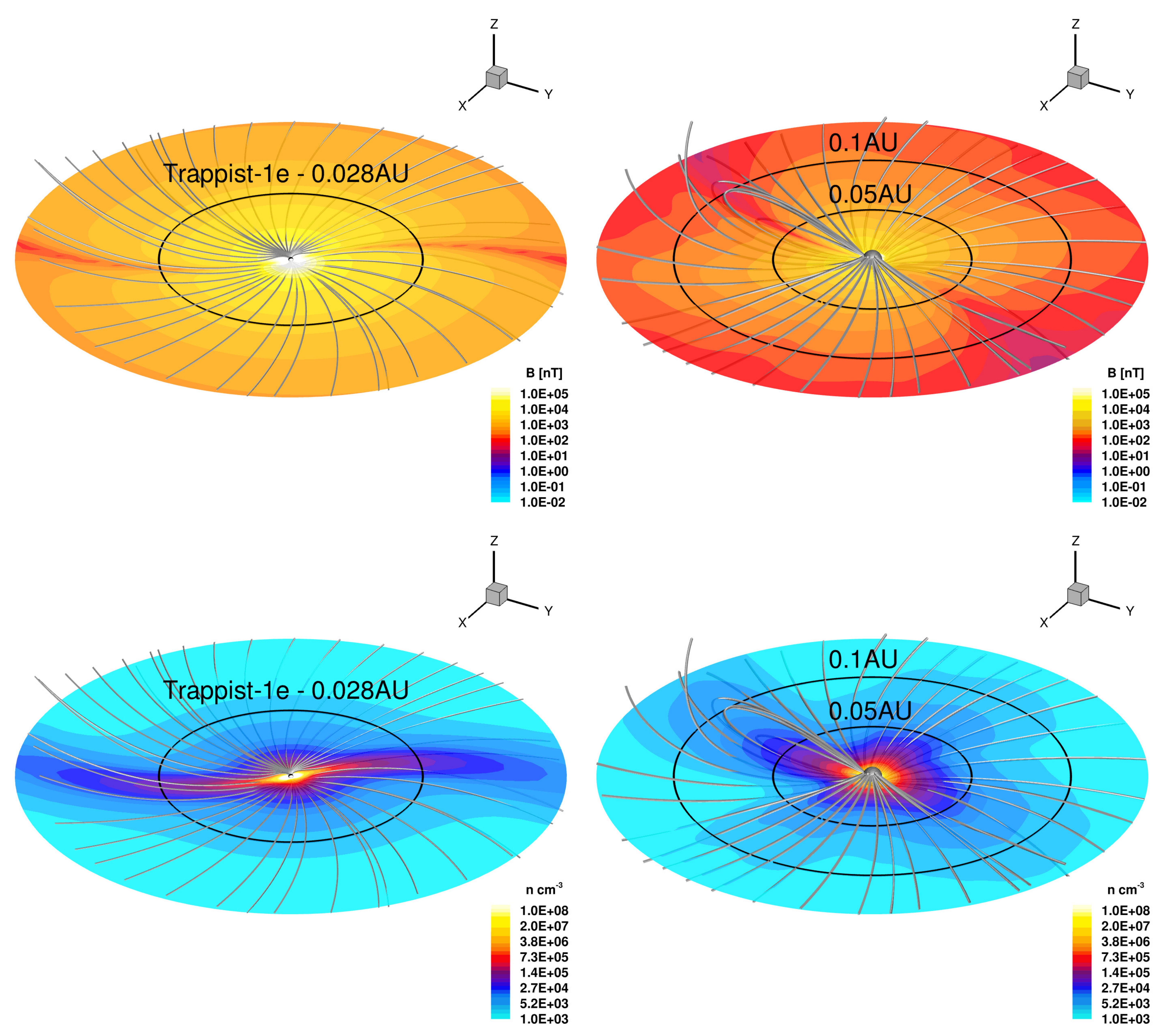}
\caption{An equatorial slices showing the solar/stellar wind solution for Trappist-1e (left column) and for solar Carrington Rotation 1916 (right column). Color contours are of magnetic field strength (top) and number density (bottom). Also shown is the orbit of Trappist-1e at 0.028AU and the hypothetical orbits at 0.05AU and 0.1AU as solid black circles. Selected magnetic field lines are shown in grey.}
\label{fig:Corona}
\end{figure*}

Figure~\ref{fig:Corona} shows the AWSoM solutions for the Sun and Trappist-1. The figure shows the distribution of the magnetic field strength and the number density over the equatorial plane in the simulation domain, along with the two hypothetical solar orbits and the orbit of the exoplanet Trappist-1e. It can be seen that the magnetic field of Trappist-1 is much stronger than that of the Sun (M-dwarf stellar fields are much stronger than those of G-type stares), leading to a much stronger variations of the field strength along the orbit of Trappist-1e. The strongest variations are seen when the planet crosses the more dense regions near the helmet streamers of the stellar corona (the largest closed magnetic loops). 


\section{Results} \label{sec:results}

\begin{figure*}[h!]
\centering
\includegraphics[width=4.in]{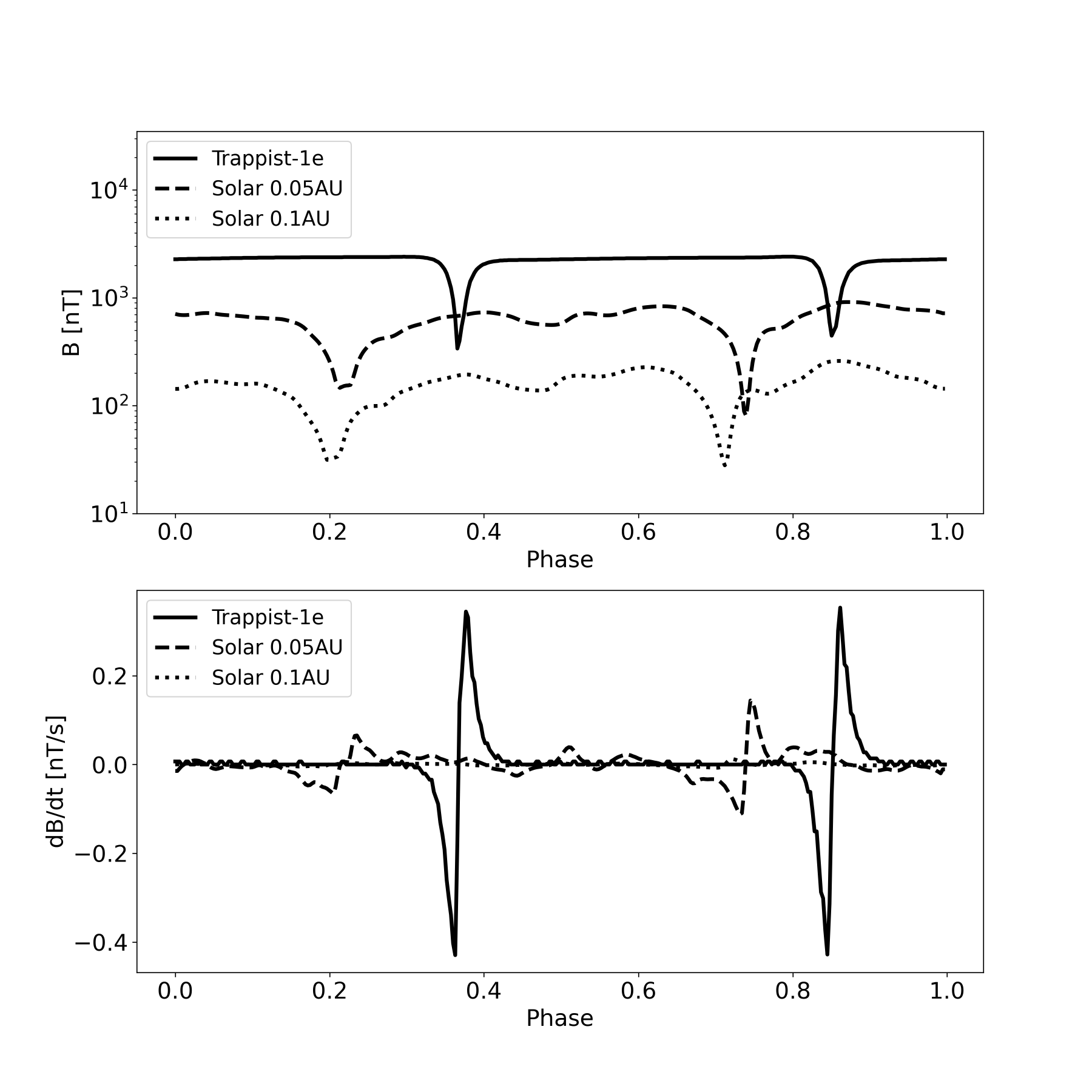}
\caption{The orbital variations of the magnetic field strength (top) and $dB/dt$ (bottom) as a function of phase for the three simulated datasets. The time derivative accounts for the specific orbital period, which is a function of the specific stellar mass and radius, and the orbital radius.}
\label{fig:B_dBdt}
\end{figure*}

Figure~\ref{fig:B_dBdt} shows the values for $B$ and $dB/dt$ for the three synthetic datasets along their associated orbits. The plots show that the the magnetic field around Trappist-1e is of the order of 1000 nT along most of the orbit, and it drops to about 600 nT during the helmet streamer crossings. Similarly, the solar hypothetical orbits of 0.05AU and 0.1AU experience IMF strength of 700-900 nT and 50-100 nT, respectively. They drop to 100 nT and 30 nT, respectively during the helmet streamer crossings. For Trappist-1e, the maximum dB/dt is $\pm0.4~nT/s$. The maximum dB/dt for the 0.05AU orbit is $\pm0.1~nT/s$ and it is smaller for the 0.1AU orbit.

The magnetic field variations here are slightly different from those presented in \cite{2018ApJ...856L..11C}. Here, we repeat the simulation with the same input conditions and parameter set, but with a slightly updated version of AWSoM, which includes a new feature that improves the physicality of the magnetic field lines in the solution. The new feature imposes the stellar wind flow and the magnetic field to be parallel to each other \citep{2022ApJ...926..102S}. This feature prevents the magnetic field lines from being "hanged" or have an inflated shape of large, closed loops. The new solution differs from that of \cite{2018ApJ...856L..11C} by having a sharper crossing of the large helmet streamers. However, the overall minimum and maximum magnitude of the magnetic field is not very different. Nevertheless, the sharper transitions lead to a stronger dB/dt and a larger heating during these times.

\begin{figure*}[h!]
\centering
\includegraphics[width=6.in]{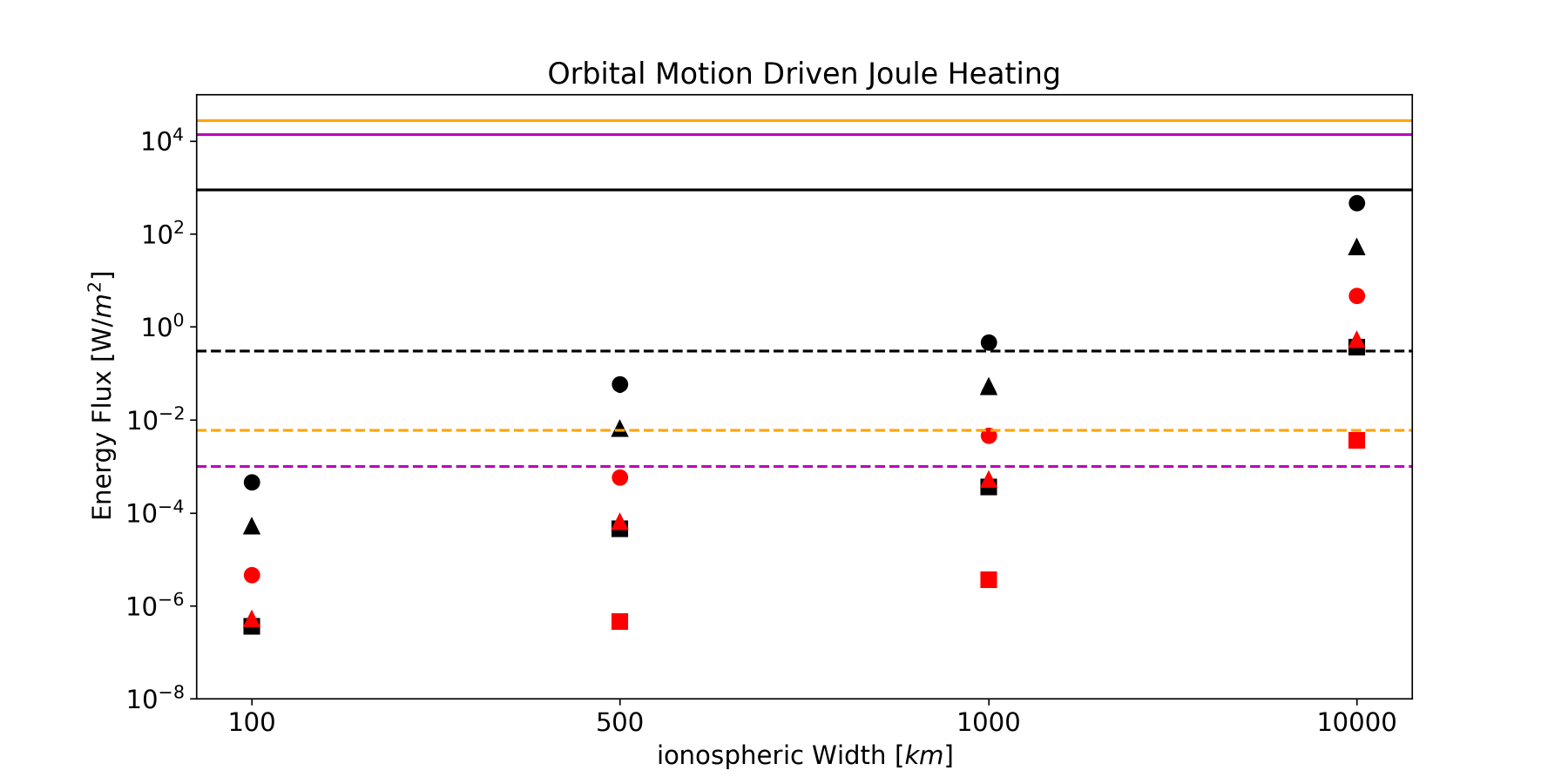}
\caption{Maximum VDJH energy flux in [$W/m^2$] as a function of ionospheric width using the maximum value of $dB/dt$ for the synthetic datasets of Trappist-1e (circles), the solar case at 0.05AU (triangles), and the solar case at 0.1AU (squares). The red markers represent the energy flux using $\sigma=0.1~[Sm^{-1}]$, while the black markers represent the energy flux using $\sigma=10~[Sm^{-1}]$. Also shown are the reference energy fluxes of the stellar/solar constant radiation (solid lines) and EUV radiation (dashed lines) for the orbits of Trappist-1e (black), 0.05AU (orange), and 0.1AU (magenta).}
\label{fig:PlanetsSigma_di}
\end{figure*}

The value of $dB/dt$ can be used to calculate the energy flux that is transported into the ionosphere. Here, we are interested in the upper limit of the VDJH. Thus, we use the maximum value of $dB/dt$ in order to calculate the maximum heating. We estimate the heating in terms of energy flux, since this parameter can be compared with other energy fluxes that may heat the upper atmosphere. Figure~\ref{fig:PlanetsSigma_di} shows the maximum values of the energy flux, $Q$, for the three synthetic datasets as a function of the ionospheric width, $d_i$. For each orbital case, we show the energy flux for ionospheric conductivities of 0.1 and 10 $Siemens$ per $meter$ ($Sm^{-1}$). These values are the lower and upper limit values of the conductivity calculated for the Earth's ionosphere \citep[e.g., recent paper by][with references therein]{2017SSRv..206..299Y} and for Neptune \citep{1989Sci...246.1478B}. Our results show that for roughly half of the cases, depending on the conductance and the ionospheric thickness, the VDJH energy flux is comparable with or larger than the EUV flux at the solar orbits of 0.05AU and 0.1AU. The highest 6 cases provide a VDJH energy flux that is comparable with or larger than the EUV flux at the orbit of Trappist-1e. The most extreme case of VDJH, with ionospheric conductivity of $10~S$, ionospheric thickness of 10,000 km, and the Trappist-1e synthetic data, reaches a value that is close to the stellar constant radiation of Trappist-1 and the orbit of Trappist-1e.

The solar constant energy flux at 0.05AU and 0.1AU is $2.74\times 10^4~Wm^{-2}$ and $1.37\times 10^4~Wm^{-2}$, respectively. The average total solar EUV irradiance at these orbits is of the order of a few $10^{-3}~Wm^{-2}$ \citep{2017SciA....3E2056M}. \cite{2017MNRAS.465L..74W,2018ApJ...856L..11C} showed that the stellar constant energy flux and the EUV energy flux at the orbit of Trappist-1e are 867 and 0.3 $Wm^{-2}$, respectively. These reference values are also shown in Figure~\ref{fig:PlanetsSigma_di}. Based on the transitions seen in Figure~\ref{fig:B_dBdt}, the peak heating events may occur a few times during the orbital period of a few days, and may have a significant integrated effect.


\section{Discussion} \label{sec:discussion}

Our results show that a rapidly changing background IMF near a short-orbit exoplanet could generate JH in the planetary ionosphere with an energy flux ranging from $0.01~Wm^{-2}$ to almost $100~Wm^{-2}$, depending of the maximum $dB/dt$, the ionospheric conductivity, and the ionospheric thickness. Comparing the reference energy fluxes to the VDJH energy flux, we find that the latter may be significant. In many of the cases, it is comparable in magnitude, or even larger than, the EUV energy flux. When considering the driving of atmospheric escape due to EUV heating, an efficiency factor is typically considered, so not all the input EUV energy is channeled to drive the escape. Therefore, VDJH is may be larger than EUV heating, even when the energy fluxes are comparable. It should be noted that here we consider a lower limit of the VDJH by considering the largest current loop, $l$, which is of the order of $r_i$. If the current can close within a smaller loop, then the VDJH will increase. For example, if instead $l$ is of the order of $d_i$, then the VDJH may be 1-2 orders of magnitude higher than the values obtained here. In this case, the VDJH energy flux may reach 1-10 percent of the stellar constant radiation at Trappist-1e (in contrast to the EUV radiation which is orders of magnitude lower). Thus, VDJH could be significant and it should be considered in the overall energy budget of atmospheres of short-orbit exoplanets. A strong VDJH may lead to a strong evaporation of the upper atmosphere (just like EUV or core-powered heating). In this scenario, the planet exhausts its own atmosphere by rapidly moving around the star in a short-orbit, sweeping by a rapidly changing, intense IMF.
 
The amount of VDJH depends on the width of the ionosphere. Short-orbit exoplanets receive an intense amount of ionizing radiation from their host star so they are expected to have an extended ionosphere \citep[e.g.,][]{2014ApJ...796...16K,2014ApJ...796...15L,2017MNRAS.469.3505W,2019AGUFM.P22B..06B}. Thus, an intense VDJH is possible in these cases. We find that VDJH is more moderate for the solar cases. However, the solar magnetic field is much weaker than that of M-dwarf stars \citep{2008MNRAS.390..545D,2008MNRAS.390..567M,2009A&A...496..787R}, which host short-orbit exoplanets that are potentially habitable (the habitable zone of M-dwarf stars is much closer to the star than that of the Sun). Therefore, the magnitude of $dB/dt$ along the planetary orbit, and the VDJH should be stronger in most potentially habitable M-dwarf exoplanets. The heating episodes occur during the transitions of the exoplanet between one plasma sector to another (the crossing of the large stellar helmet streamers). These transitions are relatively rapid, and during the rest of the orbital phases, the heating is small. To better quantify the effect investigated in this paper, a more detailed model for the upper atmosphere/ionosphere is required in order to: i) capture the integrated effect of the VDJH Vs. cooling effects over evolutionary timescales; ii) better estimate the ionospheric conductivity; iii) better estimate the extent of the ionosphere; iv) understand how atmospheric composition may play a role in the process; and v) investigate the possible drag effect on the planetary long-term orbital motion.  


\section{Summary \& Conclusions} \label{sec:summary}

We present a simple model to estimate JH of the upper atmosphere of short-orbit exoplanets. The JH is the result of a dissipation of electric current, which is driven by the rapidly varying magnetic field along the planetary orbit. We estimate the JH energy flux on the exoplanet Trappist-1e as well as similar planets orbiting the Sun in close-in orbits. We find that the JH energy flux is larger than the anticipated EUV energy flux at the planet, and it may reach a few percent of the stellar constant energy flux. Such an intense heating could drive a strong atmospheric escape and could lead to a rapid loss of the atmosphere. Thus, the rapid orbital motion of short-orbit exoplanets may exhaust a significant portion of their atmospheres over time. 

\begin{acknowledgments}

This work is supported by NASA XRP grant 80NSSC20K0840.

\end{acknowledgments}



\begin{thebibliography}{}
\expandafter\ifx\csname natexlab\endcsname\relax\def\natexlab#1{#1}\fi
\providecommand{\url}[1]{\href{#1}{#1}}
\providecommand{\dodoi}[1]{doi:~\href{http://doi.org/#1}{\nolinkurl{#1}}}
\providecommand{\doeprint}[1]{\href{http://ascl.net/#1}{\nolinkurl{http://ascl.net/#1}}}
\providecommand{\doarXiv}[1]{\href{https://arxiv.org/abs/#1}{\nolinkurl{https://arxiv.org/abs/#1}}}

\bibitem[{{Altschuler} \& {Newkirk}(1969)}]{1969SoPh....9..131A}
{Altschuler}, M.~D., \& {Newkirk}, G. 1969, Solar Physics, 9, 131,
  \dodoi{10.1007/BF00145734}

\bibitem[{{Alvarado-G{\'o}mez} {et~al.}(2019){Alvarado-G{\'o}mez}, {Garraffo},
  {Drake}, {Brown}, {Oishi}, {Moschou}, \& {Cohen}}]{2019ApJ...875L..12A}
{Alvarado-G{\'o}mez}, J.~D., {Garraffo}, C., {Drake}, J.~J., {et~al.} 2019,
  Astrophysical Journal Letters, 875, L12, \dodoi{10.3847/2041-8213/ab1489}

\bibitem[{{Alvarado-G{\'o}mez} {et~al.}(2022){Alvarado-G{\'o}mez}, {Cohen},
  {Drake}, {Fraschetti}, {Poppenhaeger}, {Garraffo}, {Chebly}, {Ilin},
  {Harbach}, \& {Kochukhov}}]{2022ApJ...928..147A}
{Alvarado-G{\'o}mez}, J.~D., {Cohen}, O., {Drake}, J.~J., {et~al.} 2022,
  Astrophysical Journal, 928, 147, \dodoi{10.3847/1538-4357/ac54b8}

\bibitem[{{Batygin} \& {Stevenson}(2010)}]{2010ApJ...714L.238B}
{Batygin}, K., \& {Stevenson}, D.~J. 2010, Astrophysical Journal Letters, 714,
  L238, \dodoi{10.1088/2041-8205/714/2/L238}

\bibitem[{{Belcher} {et~al.}(1989){Belcher}, {Bridge}, {Bagenal}, {Coppi},
  {Divers}, {Eviatar}, {Gordon}, {Lazarus}, {McNutt}, {Ogilvie}, {Richardson},
  {Siscoe}, {Sittler}, {Steinberg}, {Sullivan}, {Szabo}, {Villanueva},
  {Vasyliunas}, \& {Zhang}}]{1989Sci...246.1478B}
{Belcher}, J.~W., {Bridge}, H.~S., {Bagenal}, F., {et~al.} 1989, Science, 246,
  1478, \dodoi{10.1126/science.246.4936.1478}

\bibitem[{{Bell} {et~al.}(2019){Bell}, {Glocer}, {Cohen}, {Airapetian}, \&
  {Arras}}]{2019AGUFM.P22B..06B}
{Bell}, J.~M., {Glocer}, A., {Cohen}, O., {Airapetian}, V., \& {Arras}, P.
  2019, in AGU Fall Meeting Abstracts, Vol. 2019, P22B--06

\bibitem[{{Cohen} {et~al.}(2014){Cohen}, {Drake}, {Glocer}, {Garraffo},
  {Poppenhaeger}, {Bell}, {Ridley}, \& {Gombosi}}]{2014ApJ...790...57C}
{Cohen}, O., {Drake}, J.~J., {Glocer}, A., {et~al.} 2014, Astrophysical
  Journal, 790, 57, \dodoi{10.1088/0004-637X/790/1/57}

\bibitem[{{Cohen} {et~al.}(2020){Cohen}, {Garraffo}, {Moschou}, {Drake},
  {Alvarado-G{\'o}mez}, {Glocer}, \& {Fraschetti}}]{2020ApJ...897..101C}
{Cohen}, O., {Garraffo}, C., {Moschou}, S.-P., {et~al.} 2020, Astrophysical
  Journal, 897, 101, \dodoi{10.3847/1538-4357/ab9637}

\bibitem[{{Cohen} {et~al.}(2018){Cohen}, {Glocer}, {Garraffo}, {Drake}, \&
  {Bell}}]{2018ApJ...856L..11C}
{Cohen}, O., {Glocer}, A., {Garraffo}, C., {Drake}, J.~J., \& {Bell}, J.~M.
  2018, Astrophysical Journal Letters, 856, L11,
  \dodoi{10.3847/2041-8213/aab5b5}

\bibitem[{{Deng} {et~al.}(2011){Deng}, {Fuller-Rowell}, {Akmaev}, \&
  {Ridley}}]{2011JGRA..116.5313D}
{Deng}, Y., {Fuller-Rowell}, T.~J., {Akmaev}, R.~A., \& {Ridley}, A.~J. 2011,
  Journal of Geophysical Research (Space Physics), 116, A05313,
  \dodoi{10.1029/2010JA016019}

\bibitem[{{Donati} {et~al.}(1989){Donati}, {Semel}, \&
  {Praderie}}]{1989A&A...225..467D}
{Donati}, J.~F., {Semel}, M., \& {Praderie}, F. 1989, Astronomy and
  Astrophysics, 225, 467

\bibitem[{{Donati} {et~al.}(2008){Donati}, {Morin}, {Petit}, {Delfosse},
  {Forveille}, {Auri{\`e}re}, {Cabanac}, {Dintrans}, {Fares}, {Gastine},
  {Jardine}, {Ligni{\`e}res}, {Paletou}, {Ramirez Velez}, \&
  {Th{\'e}ado}}]{2008MNRAS.390..545D}
{Donati}, J.~F., {Morin}, J., {Petit}, P., {et~al.} 2008, MNRAS, 390, 545,
  \dodoi{10.1111/j.1365-2966.2008.13799.x}

\bibitem[{{Dong} {et~al.}(2018){Dong}, {Jin}, {Lingam}, {Airapetian}, {Ma}, \&
  {van der Holst}}]{2018PNAS..115..260D}
{Dong}, C., {Jin}, M., {Lingam}, M., {et~al.} 2018, Proceedings of the National
  Academy of Science, 115, 260, \dodoi{10.1073/pnas.1708010115}

\bibitem[{{Garraffo} {et~al.}(2022){Garraffo}, {Alvarado-G{\'o}mez}, {Cohen},
  \& {Drake}}]{2022ApJ...941L...8G}
{Garraffo}, C., {Alvarado-G{\'o}mez}, J.~D., {Cohen}, O., \& {Drake}, J.~J.
  2022, Astrophysical Journal Letters, 941, L8,
  \dodoi{10.3847/2041-8213/aca487}

\bibitem[{{Garraffo} {et~al.}(2016){Garraffo}, {Drake}, \&
  {Cohen}}]{2016ApJ...833L...4G}
{Garraffo}, C., {Drake}, J.~J., \& {Cohen}, O. 2016, Astrophysical Journal
  Letters, 833, L4, \dodoi{10.3847/2041-8205/833/1/L4}

\bibitem[{{Garraffo} {et~al.}(2017){Garraffo}, {Drake}, {Cohen},
  {Alvarado-G{\'o}mez}, \& {Moschou}}]{2017ApJ...843L..33G}
{Garraffo}, C., {Drake}, J.~J., {Cohen}, O., {Alvarado-G{\'o}mez}, J.~D., \&
  {Moschou}, S.~P. 2017, Astrophysical Journal Letters, 843, L33,
  \dodoi{10.3847/2041-8213/aa79ed}

\bibitem[{{Gillon} {et~al.}(2016){Gillon}, {Jehin}, {Lederer}, {Delrez}, {de
  Wit}, {Burdanov}, {Van Grootel}, {Burgasser}, {Triaud}, {Opitom}, {Demory},
  {Sahu}, {Bardalez Gagliuffi}, {Magain}, \& {Queloz}}]{2016Natur.533..221G}
{Gillon}, M., {Jehin}, E., {Lederer}, S.~M., {et~al.} 2016, Nature, 533, 221,
  \dodoi{10.1038/nature17448}

\bibitem[{{Ginzburg} {et~al.}(2018){Ginzburg}, {Schlichting}, \&
  {Sari}}]{Ginzburg2018}
{Ginzburg}, S., {Schlichting}, H.~E., \& {Sari}, R. 2018, MNRAS, 476, 759,
  \dodoi{10.1093/mnras/sty290}

\bibitem[{{Jackson}(1998)}]{1998clel.book.....J}
{Jackson}, J.~D. 1998, {Classical Electrodynamics, 3rd Edition} (John Wiley \&
  Sons, Inc)

\bibitem[{{Kislyakova} {et~al.}(2017){Kislyakova}, {Noack}, {Johnstone},
  {Zaitsev}, {Fossati}, {Lammer}, {Khodachenko}, {Odert}, \&
  {G{\"u}del}}]{2017NatAs...1..878K}
{Kislyakova}, K.~G., {Noack}, L., {Johnstone}, C.~P., {et~al.} 2017, Nature
  Astronomy, 1, 878, \dodoi{10.1038/s41550-017-0284-0}

\bibitem[{{Kivelson} \& {Ridley}(2008)}]{KivelsonRidley:08}
{Kivelson}, M.~G., \& {Ridley}, A.~J. 2008, Journal of Geophysical Research
  (Space Physics), 113, A05214, \dodoi{10.1029/2007JA012302}

\bibitem[{{Kivelson} \& {Russell}(1995)}]{Kivelson.Russell:95}
{Kivelson}, M.~G., \& {Russell}, C.~T. 1995, {Introduction to Space Physics},
  586

\bibitem[{{Koskinen} {et~al.}(2014){Koskinen}, {Yelle}, {Lavvas}, \& {Y-K.
  Cho}}]{2014ApJ...796...16K}
{Koskinen}, T.~T., {Yelle}, R.~V., {Lavvas}, P., \& {Y-K. Cho}, J. 2014,
  Astrophysical Journal, 796, 16, \dodoi{10.1088/0004-637X/796/1/16}

\bibitem[{{Kosovichev} \& {HMI Science Team}(2007)}]{2007AN....328..339K}
{Kosovichev}, A.~G., \& {HMI Science Team}. 2007, Astronomische Nachrichten,
  328, 339, \dodoi{10.1002/asna.200710740}

\bibitem[{{Lammer} {et~al.}(2003){Lammer}, {Selsis}, {Ribas}, {Guinan},
  {Bauer}, \& {Weiss}}]{Lammer2003}
{Lammer}, H., {Selsis}, F., {Ribas}, I., {et~al.} 2003, Astrophysical Journal
  Letters, 598, L121, \dodoi{10.1086/380815}

\bibitem[{{Lavvas} {et~al.}(2014){Lavvas}, {Koskinen}, \&
  {Yelle}}]{2014ApJ...796...15L}
{Lavvas}, P., {Koskinen}, T., \& {Yelle}, R.~V. 2014, Astrophysical Journal,
  796, 15, \dodoi{10.1088/0004-637X/796/1/15}

\bibitem[{{Morgan} \& {Taroyan}(2017)}]{2017SciA....3E2056M}
{Morgan}, H., \& {Taroyan}, Y. 2017, Science Advances, 3, e1602056,
  \dodoi{10.1126/sciadv.1602056}

\bibitem[{{Morin} {et~al.}(2008){Morin}, {Donati}, {Petit}, {Delfosse},
  {Forveille}, {Albert}, {Auri{\`e}re}, {Cabanac}, {Dintrans}, {Fares},
  {Gastine}, {Jardine}, {Ligni{\`e}res}, {Paletou}, {Ramirez Velez}, \&
  {Th{\'e}ado}}]{2008MNRAS.390..567M}
{Morin}, J., {Donati}, J.~F., {Petit}, P., {et~al.} 2008, MNRAS, 390, 567,
  \dodoi{10.1111/j.1365-2966.2008.13809.x}

\bibitem[{{Murray-Clay} {et~al.}(2009){Murray-Clay}, {Chiang}, \&
  {Murray}}]{Murray-Clay2009}
{Murray-Clay}, R.~A., {Chiang}, E.~I., \& {Murray}, N. 2009, Astrophysical
  Journal, 693, 23, \dodoi{10.1088/0004-637X/693/1/23}

\bibitem[{{Owen}(2019)}]{Owen19}
{Owen}, J.~E. 2019, Annual Review of Earth and Planetary Sciences, 47, 67,
  \dodoi{10.1146/annurev-earth-053018-060246}

\bibitem[{{Owen} \& {Jackson}(2012)}]{Owen2012}
{Owen}, J.~E., \& {Jackson}, A.~P. 2012, MNRAS, 425, 2931,
  \dodoi{10.1111/j.1365-2966.2012.21481.x}

\bibitem[{{Penz} {et~al.}(2008){Penz}, {Micela}, \& {Lammer}}]{Penz2008}
{Penz}, T., {Micela}, G., \& {Lammer}, H. 2008, Astronomy and Astrophysics,
  477, 309, \dodoi{10.1051/0004-6361:20078364}

\bibitem[{{Powell} {et~al.}(1999){Powell}, {Roe}, {Linde}, {Gombosi}, \& {De
  Zeeuw}}]{1999JCoPh.154..284P}
{Powell}, K.~G., {Roe}, P.~L., {Linde}, T.~J., {Gombosi}, T.~I., \& {De Zeeuw},
  D.~L. 1999, Journal of Computational Physics, 154, 284,
  \dodoi{10.1006/jcph.1999.6299}

\bibitem[{{Raouafi} {et~al.}(2023){Raouafi}, {Matteini}, {Squire}, {Badman},
  {Velli}, {Klein}, {Chen}, {Matthaeus}, {Szabo}, {Linton}, {Allen}, {Szalay},
  {Bruno}, {Decker}, {Akhavan-Tafti}, {Agapitov}, {Bale}, {Bandyopadhyay},
  {Battams}, {Ber{\v{c}}i{\v{c}}}, {Bourouaine}, {Bowen}, {Cattell},
  {Chandran}, {Chhiber}, {Cohen}, {D'Amicis}, {Giacalone}, {Hess}, {Howard},
  {Horbury}, {Jagarlamudi}, {Joyce}, {Kasper}, {Kinnison}, {Laker}, {Liewer},
  {Malaspina}, {Mann}, {McComas}, {Niembro-Hernandez}, {Nieves-Chinchilla},
  {Panasenco}, {Pokorn{\'y}}, {Pusack}, {Pulupa}, {Perez}, {Riley},
  {Rouillard}, {Shi}, {Stenborg}, {Tenerani}, {Verniero}, {Viall}, {Vourlidas},
  {Wood}, {Woodham}, \& {Woolley}}]{2023SSRv..219....8R}
{Raouafi}, N.~E., {Matteini}, L., {Squire}, J., {et~al.} 2023, Space Science
  Reviews, 219, 8, \dodoi{10.1007/s11214-023-00952-4}

\bibitem[{{Reiners} \& {Basri}(2009)}]{2009A&A...496..787R}
{Reiners}, A., \& {Basri}, G. 2009, Astronomy and Astrophysics, 496, 787,
  \dodoi{10.1051/0004-6361:200811450}

\bibitem[{{Rogers} \& {Showman}(2014)}]{2014ApJ...782L...4R}
{Rogers}, T.~M., \& {Showman}, A.~P. 2014, Astrophysical Journal Letters, 782,
  L4, \dodoi{10.1088/2041-8205/782/1/L4}

\bibitem[{{Russell}(2009)}]{SpacePhysics16}
{Russell}, C.~T. 2009, {Space Physics: An Introduction}

\bibitem[{Salz {et~al.}(2016)Salz, Schneider, Czesla, \&
  Schmitt}]{salz2016energy}
Salz, M., Schneider, P., Czesla, S., \& Schmitt, J. 2016, Astronomy and
  Astrophysics, 585, L2

\bibitem[{{Scherrer} {et~al.}(1991){Scherrer}, {Hoeksema}, \&
  {Bush}}]{1991AdSpR..11d.113S}
{Scherrer}, P.~H., {Hoeksema}, J.~T., \& {Bush}, R.~I. 1991, Advances in Space
  Research, 11, 113, \dodoi{10.1016/0273-1177(91)90446-Q}

\bibitem[{{Schneiter} {et~al.}(2007){Schneiter}, {Vel{\'a}zquez}, {Esquivel},
  {Raga}, \& {Blanco-Cano}}]{Schneiter2007}
{Schneiter}, E.~M., {Vel{\'a}zquez}, P.~F., {Esquivel}, A., {Raga}, A.~C., \&
  {Blanco-Cano}, X. 2007, Astrophysical Journal Letters, 671, L57,
  \dodoi{10.1086/524945}

\bibitem[{{Schrijver} \& {Siscoe}(2009)}]{Heliophysics09}
{Schrijver}, C.~J., \& {Siscoe}, G.~L. 2009, {Heliophysics: Plasma Physics of
  the Local Cosmos}

\bibitem[{{Showman} {et~al.}(2010){Showman}, {Cho}, \&
  {Menou}}]{2010exop.book..471S}
{Showman}, A.~P., {Cho}, J.~Y.~K., \& {Menou}, K. 2010, in Exoplanets, ed.
  S.~{Seager}, 471--516, \dodoi{10.48550/arXiv.0911.3170}

\bibitem[{{Sokolov} {et~al.}(2022){Sokolov}, {Zhao}, \&
  {Gombosi}}]{2022ApJ...926..102S}
{Sokolov}, I.~V., {Zhao}, L., \& {Gombosi}, T.~I. 2022, Astrophysical Journal,
  926, 102, \dodoi{10.3847/1538-4357/ac400f}

\bibitem[{{Sokolov} {et~al.}(2021){Sokolov}, {Holst}, {Manchester}, {Su
  Ozturk}, {Szente}, {Taktakishvili}, {T{\'o}th}, {Jin}, \&
  {Gombosi}}]{2021ApJ...908..172S}
{Sokolov}, I.~V., {Holst}, B. v.~d., {Manchester}, W.~B., {et~al.} 2021,
  Astrophysical Journal, 908, 172, \dodoi{10.3847/1538-4357/abc000}

\bibitem[{Tian {et~al.}(2005)Tian, Toon, Pavlov, \& Sterck}]{Tian2005extra}
Tian, F., Toon, O.~B., Pavlov, A.~A., \& Sterck, H.~D. 2005, Astrophysical
  Journal, 621, 1049.
\newblock \url{http://stacks.iop.org/0004-637X/621/i=2/a=1049}

\bibitem[{{T{\'o}th} {et~al.}(2012){T{\'o}th}, {van der Holst}, {Sokolov}, {De
  Zeeuw}, {Gombosi}, {Fang}, {Manchester}, {Meng}, {Najib}, {Powell}, {Stout},
  {Glocer}, {Ma}, \& {Opher}}]{2012JCoPh.231..870T}
{T{\'o}th}, G., {van der Holst}, B., {Sokolov}, I.~V., {et~al.} 2012, Journal
  of Computational Physics, 231, 870, \dodoi{10.1016/j.jcp.2011.02.006}

\bibitem[{{Tripathi} {et~al.}(2015){Tripathi}, {Kratter}, {Murray-Clay}, \&
  {Krumholz}}]{Tripathi2015}
{Tripathi}, A., {Kratter}, K.~M., {Murray-Clay}, R.~A., \& {Krumholz}, M.~R.
  2015, Astrophysical Journal, 808, 173, \dodoi{10.1088/0004-637X/808/2/173}

\bibitem[{{Valencia} {et~al.}(2018){Valencia}, {Tan}, \&
  {Zajac}}]{2018ApJ...857..106V}
{Valencia}, D., {Tan}, V. Y.~Y., \& {Zajac}, Z. 2018, Astrophysical Journal,
  857, 106, \dodoi{10.3847/1538-4357/aab767}

\bibitem[{{van der Holst} {et~al.}(2014){van der Holst}, {Sokolov}, {Meng},
  {Jin}, {Manchester}, {T{\'o}th}, \& {Gombosi}}]{2014ApJ...782...81V}
{van der Holst}, B., {Sokolov}, I.~V., {Meng}, X., {et~al.} 2014, Astrophysical
  Journal, 782, 81, \dodoi{10.1088/0004-637X/782/2/81}

\bibitem[{{Vidotto} {et~al.}(2018){Vidotto}, {Lichtenegger}, {Fossati},
  {Folsom}, {Wood}, {Murthy}, {Petit}, {Sreejith}, \&
  {Valyavin}}]{2018MNRAS.481.5296V}
{Vidotto}, A.~A., {Lichtenegger}, H., {Fossati}, L., {et~al.} 2018, MNRAS, 481,
  5296, \dodoi{10.1093/mnras/sty2130}

\bibitem[{Watson {et~al.}(1981)Watson, Donahue, \&
  Walker}]{watson_dynamics_1981}
Watson, A.~J., Donahue, T.~M., \& Walker, J.~C. 1981, Icarus, 48, 150,
  \dodoi{10.1016/0019-1035(81)90101-9}

\bibitem[{{Weber} {et~al.}(2017){Weber}, {Lammer}, {Shaikhislamov}, {Chadney},
  {Khodachenko}, {Grie{\ss}meier}, {Rucker}, {Vocks}, {Macher}, {Odert}, \&
  {Kislyakova}}]{2017MNRAS.469.3505W}
{Weber}, C., {Lammer}, H., {Shaikhislamov}, I.~F., {et~al.} 2017, MNRAS, 469,
  3505, \dodoi{10.1093/mnras/stx1099}

\bibitem[{{Wheatley} {et~al.}(2017){Wheatley}, {Louden}, {Bourrier},
  {Ehrenreich}, \& {Gillon}}]{2017MNRAS.465L..74W}
{Wheatley}, P.~J., {Louden}, T., {Bourrier}, V., {Ehrenreich}, D., \& {Gillon},
  M. 2017, MNRAS, 465, L74, \dodoi{10.1093/mnrasl/slw192}

\bibitem[{{Wood} {et~al.}(2021){Wood}, {M{\"u}ller}, {Redfield}, {Konow},
  {Vannier}, {Linsky}, {Youngblood}, {Vidotto}, {Jardine},
  {Alvarado-G{\'o}mez}, \& {Drake}}]{2021ApJ...915...37W}
{Wood}, B.~E., {M{\"u}ller}, H.-R., {Redfield}, S., {et~al.} 2021,
  Astrophysical Journal, 915, 37, \dodoi{10.3847/1538-4357/abfda5}

\bibitem[{{Yamazaki} \& {Maute}(2017)}]{2017SSRv..206..299Y}
{Yamazaki}, Y., \& {Maute}, A. 2017, Space Science Reviews, 206, 299,
  \dodoi{10.1007/s11214-016-0282-z}

\end{thebibliography}


\end{document}